\documentclass[pdftex,twocolumn,epjc3]{svjour3}          

\RequirePackage[T1]{fontenc}

\smartqed  
\usepackage{amsmath}
\usepackage{amssymb}
\usepackage{amsfonts}
\RequirePackage{graphicx}
\usepackage{subcaption} 
\usepackage{caption}    

\RequirePackage{mathptmx}      
\RequirePackage{flushend}
\RequirePackage[numbers,sort&compress]{natbib}
\RequirePackage[colorlinks,citecolor=blue,urlcolor=blue,linkcolor=blue]{hyperref}
\usepackage{subcaption}
\usepackage{booktabs}  
\usepackage{adjustbox} 

\journalname{Eur. Phys. J. C}

\begin{document}

\title{Topological Landscapes of the BSM Higgs Sector}


\author{Jyotiranjan Beuria\thanksref{addr1}}

\thankstext{e1}{e-mail: jyotiranjan.beuria@gmail.com}

\institute{IKS Research Centre, ISS Delhi, Delhi, India\label{addr1}
}

\date{}

\maketitle

\begin{abstract}
We explore the structure of the parameter space in the Singlet Scalar Dark Matter (SSDM) model and the Next-to-Two Higgs Doublet Model (N2HDM) with $\tan\beta = 5$ and $\tan\beta = 45$. Parameter points are classified as allowed or excluded based on compatibility with the Higgs observation constraints. Using a combined framework of Topological Data Analysis (TDA), Uniform Manifold Approximation and Projection (UMAP), and Linear Discriminant Analysis (LDA), we characterize the global geometry and topology of these high-dimensional landscapes. Our findings reveal that the SSDM and the N2HDM Type~I model exhibit finely tuned islands of collider viability. In the case of N2HDM Type~I, we also find that increasing $\tan\beta$ leads to greater topological fragmentation and higher Betti number persistence, indicating enhanced structural complexity. In contrast, the given choice of parameters excludes the entire N2HDM Type II parameter space based on current Higgs measurements. The topological properties serve as important quantitative descriptors for the phenomenological viability of the BSM frameworks. We leverage the above mentioned topological features to train machine learning models for faster characterization of the BSM Higgs sector into allowed and excluded parameter regions.
\end{abstract}

\section{Introduction}

The discovery of the Higgs boson at the LHC has completed the Standard Model (SM), yet the structure of the Higgs sector remains an open question.  
Extensions of the scalar sector are theoretically well motivated and arise in 
a variety of contexts, including explanations of dark matter \cite{baek2014invisible,arcadi2020dark,dutta2024dark}, electroweak 
baryogenesis \cite{morrissey2012electroweak}, naturalness \cite{barbieri2006improved,kang2012naturalness}, and flavor physics \cite{ manohar2006flavor,crivellin2013flavor}.  
Such beyond the Standard Model (BSM) scenarios typically introduce additional 
degrees of freedom and new scalar interactions, enlarging the parameter space 
far beyond that of the SM.

Since the BSM parameter space is inherently high dimensional and has non-linear dependencies \cite{baruah2024probing}, its characterisation in light of phenomenological bounds is computationally demanding. The global organization in these high-dimensional parameter spaces 
is essential for assessing the viability of different BSM models under current 
and future collider constraints.  
Traditionally, phenomenological studies proceed by scanning parameters over 
pre-defined ranges, applying collider and theoretical constraints 
(e.g., vacuum stability, perturbativity, and collider limits), 
and then visualizing the results through scatter plots or two-dimensional 
projections.  
While effective for highlighting specific correlations, such approaches are 
inherently limited: they obscure the global geometry of the viable regions and 
offer little insight into the connectivity or fragmentation of allowed domains \cite{beuria2023persistent,beuria2024intrinsic}.  
As experimental data tighten, it becomes increasingly important not only to 
identify viable parameter points, but also to efficiently characterize the shape and 
topology of the regions they occupy.  

Recent developments in machine learning (ML) have greatly accelerated such analyses by employing techniques like Bayesian neural networks, active learning, and simulation-based inference. These methods enable efficient parameter scans, emulate complex theoretical predictions, and identify viable model regions with far fewer evaluations. Some recent implementations include ML-accelerated BSM interpretations of collider data~\cite{bertone2019accelerating,beuria2025probing}, Bayesian neural networks for SUSY predictions~\cite{kronheim2021bayesian,abreu2023exploring}, active learning parameter scans~\cite{goodsell2023active,goodsell2024bsmart}, and neural network–aided simulation-based inference~\cite{chatterjee2025exploring}.

Recent advances in data science provide powerful new tools for characterizing such parameter spaces. Machine learning techniques allow us to identify nonlinear structures and decision boundaries that are invisible in conventional approaches. Beyond this, topological data analysis (TDA) \cite{carlsson2020topological,chazal2021introduction}, and in particular persistent homology, provides a rigorous framework for quantifying 
global features of parameter spaces.  
By computing Betti numbers as a function of a filtration scale, persistent homology 
reveals whether allowed regions form a single connected domain, fragment into 
multiple islands, or contain higher-dimensional loop structures.  
These topological invariants complement traditional statistical measures and machine learning frameworks, offering a deeper geometric perspective on model viability.  

In this work, we apply such methods to two representative BSM Higgs scenarios.  
The first is the Singlet Scalar Dark Matter (SSDM) model \cite{guo2010real,cline2013update,gambit2017status}, a minimal Higgs-portal extension that introduces a $\mathbb{Z}_2$-stabilized 
scalar singlet as a dark matter candidate.  
The second is the Next-to-2HDM (N2HDM) \cite{muhlleitner2017n2hdm,biekotter2022possible}, which augments the 
two-Higgs-doublet model with an additional real singlet scalar. We consider two of its variants, namely, the so-called Type~I and Type~II, defined in line with the Yukawa structures found in 2HDM models. These models span complementary directions: the SSDM provides the simplest  
portal to dark matter, while the N2HDM realizes a multi-scalar Higgs sector 
with rich mixing patterns. In the N2HDM, we consider two benchmark values of the Higgs doublet vacuum expectation values (VEV) ratio, $\tan\beta = 5$ and $\tan\beta = 45$, to study how the Higgs-sector phenomenology changes across different coupling regimes.

Our strategy is as follows.  
We generate large parameter scans ($O(10^6)$) for each model, label points as allowed 
or excluded according to collider constraints for an SM-like Higgs. We analyze the resulting global
landscapes using Betti number distributions from persistent homology considerations \cite{beuria2023persistent, beuria2024intrinsic, beuria2025probing}. We find that allowed regions form a topologically rich landscape compared to the excluded parameter space. We further study the structure of the BSM Higgs parameter space using supervised Uniform Manifold Approximation and Projection (UMAP) \cite{mcinnes2018umap,ghojogh2021uniform}, and 
Linear Discriminant Analysis (LDA) \cite{balakrishnama1998linear,zhao2024linear}. Finally, we develop a machine learning model for faster characterization of the BSM parameter space into allowed and excluded regions by considering limits on observed final states and goodness of fit.  
 
The organisation of this paper is as follows.  
In section \ref{sec:models}, we introduce the SSDM and N2HDM models and define the scanning 
parameters that govern their Higgs-sector phenomenology.  
In section \ref{sec:framework}, we describe our analysis methodology involving TDA, UMAP and LDA. In section \ref{sec:ml}, we present a machine learning framework to classify phenomenologically viable and nonviable parameter regions. 
Section \ref{sec:results} presents the results for each model and compares their 
topological landscapes.  
Finally, we conclude in section \ref{sec:conclusion}.

\section{Models and Parameters}
\label{sec:models}
In this work we investigate two well-motivated extensions of the Standard Model (SM) Higgs sector: the Singlet Scalar Dark Matter (SSDM) model and the Next-to-2HDM (N2HDM), including its Type~I and Type~II variants.  Both frameworks enlarge the scalar sector and modify the Higgs phenomenology, 
leading to distinct patterns of allowed and excluded regions under the current collider and theoretical constraints. The aim of our study is to characterize the global structure of these parameter spaces, with particular emphasis on the regions consistent with the observed properties of the SM Higgs. We discuss the two models below before diving into the scan and characterization of their respective parameter space.

\subsection{Singlet Scalar Dark Matter (SSDM)}

The SSDM extends the SM by adding a real scalar singlet $S$, which is stabilized 
by a discrete $\mathbb{Z}_2$ symmetry ($S \rightarrow -S$) and thus serves as a viable dark matter candidate.  
The scalar potential takes the form
\begin{equation}
V(H,S) = \mu_H^2 |H|^2 + \lambda_H |H|^4 
+ \frac{1}{2} m_S^2 S^2 + \frac{1}{4} \lambda_S S^4
+ \frac{1}{2} \lambda_{SH} |H|^2 S^2 ,
\end{equation}
where $H$ is the SM Higgs doublet.  
The coupling $\lambda_{SH}$ controls the Higgs–portal interaction and directly 
governs Higgs production and decay rates in the presence of the singlet.  
The quartic self-coupling $\lambda_S$ and the singlet mass parameter $M_S^2$ 
determine the mass and stability of the dark matter state.  Thus, the relevant free parameters that shape Higgs-sector phenomenology are: $\lambda_S, \; \lambda_{SH}, \; m_S^2$. We vary these three parameters in our scan to explore how observed properties of the SM Higgs partition the SSDM parameter space into allowed and excluded regions.

\subsection{Next-to-2HDM (N2HDM)}

The Next-to-2HDM (N2HDM) is a well-motivated extension of the 
two-Higgs-doublet model (2HDM) by the addition of a real scalar singlet $S$.  
Its scalar field content thus consists of two $SU(2)_L$ doublets, $\Phi_1$ and $\Phi_2$, 
and one real gauge singlet, $S$.  
After electroweak symmetry breaking (EWSB), this leads to an enlarged scalar spectrum: 
three neutral CP-even Higgs bosons, one neutral CP-odd Higgs boson, and a pair of charged Higgs bosons.  
The extended CP-even sector arises from the mixing of the two doublets with the singlet, 
providing a richer Higgs phenomenology than the minimal 2HDM, and allowing for signatures 
such as singlet admixtures in the observed $125$~GeV Higgs boson or additional 
light/heavy CP-even Higgs states.

The most general renormalizable and CP-conserving scalar potential of the N2HDM, 
consistent with a softly broken $\mathbb{Z}_2$ symmetry, is given by
\begin{align}
V(\Phi_1, \Phi_2, S) &= 
m_{11}^2 \, \Phi_1^\dagger \Phi_1
+ m_{22}^2 \, \Phi_2^\dagger \Phi_2
- m_{12}^2 \, (\Phi_1^\dagger \Phi_2 + \Phi_2^\dagger \Phi_1)
\nonumber \\[4pt]
&+ \frac{\lambda_1}{2} (\Phi_1^\dagger \Phi_1)^2
+ \frac{\lambda_2}{2} (\Phi_2^\dagger \Phi_2)^2
+ \lambda_3 (\Phi_1^\dagger \Phi_1)(\Phi_2^\dagger \Phi_2)
\nonumber \\[4pt]
&+ \lambda_4 (\Phi_1^\dagger \Phi_2)(\Phi_2^\dagger \Phi_1)
+ \frac{\lambda_5}{2}\big[ (\Phi_1^\dagger \Phi_2)^2 
+ (\Phi_2^\dagger \Phi_1)^2 \big]
\nonumber \\[6pt]
&+ \frac{1}{2} m_S^2 \, S^2
+ \frac{\lambda_6}{8} S^4
+ \frac{\lambda_7}{2} (\Phi_1^\dagger \Phi_1) S^2
+ \frac{\lambda_8}{2} (\Phi_2^\dagger \Phi_2) S^2 .
\label{eq:pot-n2hdm}
\end{align}

Here, $m_{11}^2$, $m_{22}^2$, $m_{12}^2$, and $m_S^2$ are mass parameters, 
while $\lambda_{1\text{--}8}$ are dimensionless quartic couplings.  
The doublet interactions ($\lambda_{1\text{--}5}$) reduce to the ordinary 2HDM potential, 
while $\lambda_6$ controls the singlet self-interaction, 
and $\lambda_{7,8}$ act as ``portal'' couplings that mediate mixing between the singlet and the doublets.  
The singlet admixture modifies the couplings of the physical Higgs bosons to fermions and gauge bosons, 
making precision Higgs measurements a sensitive probe of this model.

For Higgs phenomenology, particularly, for the singlet sector, the most relevant free parameters in our analysis are $\lambda_6, \, \lambda_7, \, \lambda_8, \,\text{and}\, m_S^2$, which govern the singlet-doublet mixing and the additional CP-even mass eigenstates. Now, we discuss the variation of the model with the so-called Type~I vs Type~II Yukawa structure.

\subsubsection{Yukawa sector and Type~I vs Type~II}  
A crucial ingredient of the N2HDM is its Yukawa structure.  
To forbid tree-level flavor-changing neutral currents, a discrete $\mathbb{Z}_2$ symmetry 
is imposed on the doublets, leading to several possible Yukawa assignments.  
We focus on the two canonical cases:

\paragraph{Type~I.} 
All fermions couple exclusively to the second Higgs doublet $\Phi_2$, 
while $\Phi_1$ does not couple to fermions (fermiophobic).  
Consequently, the fermionic couplings of the physical Higgs states are universally scaled by 
$\sin\alpha/\sin\beta$ or $\cos\alpha/\sin\beta$, where $\alpha$ is the CP-even mixing angle and
\begin{equation}
\tan\beta = \frac{v_2}{v_1}
\end{equation}
denotes the ratio of the vacuum expectation values of the two doublets.  
At large $\tan\beta$, the couplings to fermions become suppressed, which weakens 
direct collider limits and allows more freedom in the viable Higgs parameter space. This opens up the possibility of singlet-enriched Higgs states that evade current bounds.


\paragraph{Type~II.} 
In contrast, the Type~II assignment splits the Yukawa couplings: 
up-type quarks couple to $\Phi_2$, while down-type quarks and charged leptons couple to $\Phi_1$.  
This structure mirrors that of the MSSM Higgs sector.  
As a result, at large $\tan\beta$ the couplings to down-type fermions 
($b$ quarks, $\tau$ leptons) are enhanced, while couplings to up-type fermions are suppressed.  
This has two major consequences:  
(i) strong collider bounds from searches in $b\bar{b}$ and $\tau^+\tau^-$ final states,  
(ii) stringent flavor physics constraints, e.g.\ from $B$-meson decays.  
Hence, much of the Type~II parameter space is tightly constrained or excluded 
by current Higgs precision measurements.

At moderate or low $\tan\beta$ values, the situation becomes even more severe due to enhanced top-quark couplings. This spoils the fit to Higgs signal strengths in the $\gamma\gamma$, $ZZ^*$, and $WW^*$ channels. In addition, limits from heavy Higgs searches and flavor constraints such as $B\to X_s\gamma$ further tighten the bounds.

In this work, we systematically scan the parameter space 
$(\lambda_6, \lambda_7, \lambda_8, m_S^2)$ 
for both Type~I and Type~II, considering benchmark values 
$\tan\beta=5$ (moderate) and $\tan\beta=45$ (large).  Consequently, our scans show that for both low and high $\tan\beta$ along with other chosen representative parameter values, the entire Type~II parameter space is ruled out due to Higgs precision data and other collider limits. This is quite consistent with some recent works, wherein a highly constrained parameter space for such a model has already been explored \cite{biekotter2022possible,heinemeyer2022phenomenology}. Thus, we do not discuss its topological landscapes in our subsequent analysis.

\subsection{Scan over Parameter space}

For our numerical analysis, we implement both the Singlet Scalar Dark Matter (SSDM) 
model and the Next-to-2HDM (N2HDM) of Type~I and Type~II within the \texttt{SARAH--SPheno} framework.  
The model files are generated with \texttt{SARAH v4.15.3}~\cite{Staub:2013tta}, 
and spectrum generation, decay widths, and Higgs observables are computed using 
\texttt{SPheno v4.0.5}~\cite{Porod:2011nf}.  
The output is interfaced through \texttt{BSMArt v1.3} \cite{goodsell2024bsmart}, which provides a streamlined workflow to \texttt{HiggsTools v1.2}~\cite{Bahl:2022igd} for applying constraints from numerous Higgs-related measurements at colliders. The latter combines the constraints from \texttt{HiggsBounds v5} \cite{Bahl:2021yhk} and \texttt{HiggsSignals v2} \cite{Bechtle:2020uwn}, allowing each sampled parameter point to be tested 
against current collider and precision Higgs data.  

The scans are performed over the most relevant quartic couplings and 
mass parameters governing the Higgs-singlet mixing in extended 
scalar sector.  
For the Singlet Scalar Dark Matter (SSDM) model, we scan over the parameters:
\begin{equation}
\begin{aligned}
\lambda_S \in [-1,\,1], \quad
\lambda_{SH} \in [-1,\,1],\quad
|m_S^2| < 5000~\text{GeV}^2 \,.
\end{aligned}
\end{equation}
This choice captures the key directions that determine both the Higgs sector phenomenology and the viability of the dark matter candidate. However, we refrain from dark matter related constraints and focus exclusively on collider bounds. 

For the Next-to-2HDM (N2HDM), we vary the portal couplings 
\begin{equation}
\begin{aligned}
\lambda_6 \in [-1,\,1], \quad
\lambda_7 \in [-1,\,1], \quad
\lambda_8 \in [-1,\,1], \\
|m_S^2| < 5000~\text{GeV}^2 \, .
\end{aligned}
\end{equation}
while keeping the remaining doublet quartics fixed to representative values. In particular, we choose  
$\lambda_1 = 0.65$, $\lambda_2 =0.15$, $\lambda_3 = -0.5$, 
$\lambda_4 = 0.35$, $\lambda_5 = 0.03$, and  $m_{12}=-4000$ GeV.
In both Type~I and Type~II realizations, we consider benchmark values $\tan\beta = 5$ and $\tan\beta = 45$ in order to contrast the low- and high-$\tan\beta$ regimes. We only consider the parameter points giving rise to at least one neutral scalar of mass, $125 \pm 3$ GeV. Thus, each scanned point is categorized as ``allowed'' or ``excluded'' 
depending on the outcome of the \texttt{HiggsTools} evaluation, which encodes the global compatibility with Higgs precision measurements.  The resulting datasets, typically comprising $O(10^6)$ points for each model, form the basis for the machine learning and topological analyses presented in this work.


\section{Clustering Framework for Parameter Space Landscapes}
\label{sec:framework}
In order to uncover the hidden structure of the BSM Higgs parameter space, we 
employ a multi-pronged framework that combines 
(i) Topological Data Analysis (TDA), 
(ii) manifold learning via UMAP, and 
(iii) Linear Discriminant Analysis (LDA).  
Each of these tools provides complementary information: TDA captures global 
topological features such as connected components and loops, UMAP embeds 
high-dimensional spaces into two dimensions while preserving local geometry, 
and LDA yields maximally discriminating projections between allowed and 
excluded regions.  

\subsection{Topological Data Analysis and Betti number Profiles}

Topological data analysis (TDA) characterizes the ``shape'' of a dataset by 
computing invariants that are robust to deformations.  
Given a set of parameter points $\mathcal{X} = \{x_i\}$ with $x_i \in \mathbb{R}^d$, 
one constructs a nested family of simplicial complexes 
$K_\epsilon(\mathcal{X})$ parameterized by a scale $\epsilon$, 
for example using Vietoris--Rips or $\alpha$-complexes.  
The evolution of homology groups $H_k(K_\epsilon)$ as $\epsilon$ increases 
gives rise to persistent homology, whose ranks 
\begin{equation}
\beta_k(\epsilon) = \mathrm{rank}\, H_k(K_\epsilon)
\end{equation}
are the Betti numbers.  
The $\beta_0$ curve describes the number of connected components, 
while $\beta_1$ corresponds to the number of independent loops in the data.  
Betti curves $\beta_k(\epsilon)$ thus provide a robust summary of the 
global topological structure of allowed vs.\ excluded regions.  

\subsection{Manifold Learning with UMAP}

While topological data analysis (TDA) captures the global shape of data, manifold learning aims to preserve local geometric relationships when 
embedding high-dimensional data into lower dimensions.  
Uniform Manifold Approximation and Projection (UMAP)~\cite{mcinnes2018umap} 
is a manifold learning algorithm that constructs a weighted 
$k$-nearest-neighbor (kNN) graph to approximate the dataset’s 
fuzzy topological structure.  
Given a dataset $\{x_i\}$ with pairwise distances $d(x_i,x_j)$, 
UMAP defines directed edge weights
\begin{equation}
\tilde{\mu}_{ij} = \exp\!\left[-\frac{\max(0,\, d(x_i,x_j) - \rho_i)}{\sigma_i}\right],
\end{equation}
where $\rho_i$ sets a local connectivity threshold ensuring that each point 
is connected to at least one neighbor with unit strength, and 
$\sigma_i$ is chosen such that the effective number of neighbors is fixed 
via $\sum_j \exp[-(d(x_i,x_j) - \rho_i)/\sigma_i] = \log_2(k)$.  
The resulting directed graph is symmetrized using a fuzzy union:
\begin{equation}
\mu_{ij} = \tilde{\mu}_{ij} + \tilde{\mu}_{ji} - \tilde{\mu}_{ij}\tilde{\mu}_{ji},
\end{equation}
yielding a single undirected weighted graph that represents 
the fuzzy simplicial set structure of the data manifold.  

A low-dimensional embedding $\{y_i\}$ is then obtained by minimizing the 
cross-entropy between the high- and low-dimensional fuzzy sets:
\begin{equation}
\mathcal{L}_{\mathrm{UMAP}} 
= - \sum_{i<j} \left[
\mu_{ij} \log \nu_{ij} + (1-\mu_{ij}) \log (1-\nu_{ij})
\right],
\end{equation}
where $\nu_{ij}$ are analogous membership strengths computed from the 
low-dimensional coordinates $\{y_i\}$.  
This optimization encourages points that are close in the high-dimensional space to remain close in the embedding, while permitting large-scale 
rearrangements that clarify global and cluster-level structure.  

In our application, we employ a supervised form of UMAP in which class labels guide the neighborhood graph: connections between points of the same class are reinforced, whereas edges across classes are down-weighted. This label-driven adjustment shapes the low-dimensional optimization to emphasize boundaries between allowed and excluded regions of parameter space, improving class separability in the embedding.

\subsection{Linear Discriminant Analysis}

Linear discriminant analysis (LDA) provides a complementary, linear projection 
technique that maximizes class separability.  
Given two classes of points, LDA constructs a projection vector $w$ such that 
the ratio of between-class to within-class variance is maximized:  
\begin{equation}
J(w) = \frac{w^\top S_B w}{w^\top S_W w} ,
\end{equation}
where $S_B$ is the between-class scatter matrix and $S_W$ the within-class 
scatter matrix.  
For binary classification, this reduces to
\begin{equation}
w \propto S_W^{-1} (\mu_1 - \mu_0) ,
\end{equation}
with $\mu_{0,1}$ denoting the class means.  
Projecting data onto $w$ yields a one-dimensional axis where allowed and 
excluded regions are most cleanly separated, providing a useful linear 
benchmark to compare with the nonlinear structures uncovered by TDA and UMAP. 

Together, these methods furnish a comprehensive view of the BSM parameter landscape: 
TDA captures global topological organization, UMAP reveals nonlinear manifold geometry, 
and LDA highlights the dominant linear discriminant directions.

\section{ML Classification and Regression in BSM Parameter space}
\label{sec:ml}
We employ a multi-label neural network framework to jointly learn binary collider-viability 
classification and $\chi^2$ regression across BSM parameter spaces. For the SSDM, each sample corresponds to a parameter-space point labeled by
$(\lambda_S,\, \lambda_{SH},\, m_{S}^2)$. Its collider viability status and global goodness-of-fit $\chi^2$ value are derived from the \texttt{HiggsTools} package as discussed earlier. For the N2HDM models, the feature space is four-dimensional, labeled by $(\lambda_6, \lambda_7, \lambda_8, m_S^2)$.

All input features are standardized using \texttt{StandardScaler} to a zero mean and unit variance. 
The shared representation network consists of two fully connected (dense) layers with 
64 and 32 neurons, respectively, each employing \texttt{ReLU} activations and a 
dropout rate of 0.2 to mitigate overfitting. The network bifurcates into two output heads: 
(i) a classification head with a single neuron and a sigmoid activation, 
predicting the binary collider-viability label $y_{\mathrm{class}} \in \{0, 1\}$ 
(ii) a regression head with a single linear neuron predicting the 
log-transformed goodness-of-fit value, $y_{\mathrm{reg}} = \log(1+\chi^2)$. 
This dual-head structure allows simultaneous learning of discrete and continuous 
targets, facilitating joint optimization of collider constraints and fit performance.

The model is trained using the \texttt{Adam} optimizer 
with a learning rate of $10^{-3}$ and a composite loss function:
\begin{equation}
\mathcal{L}_{\mathrm{total}} 
= \mathcal{L}_{\mathrm{class}} + \alpha \, \mathcal{L}_{\mathrm{reg}},
\end{equation}
where $\mathcal{L}_{\mathrm{class}}$ is the binary cross-entropy loss,
$\mathcal{L}_{\mathrm{reg}}$ is the Huber loss for robust regression, 
and $\alpha = 0.5$ controls the relative weighting. 
The choice of the Huber loss function is motivated by its 
robustness to large residuals, which are common for parameter-space points 
with extreme $\chi^2$ values. 

Training is performed for 30 epochs with a batch size of 256, using an 80-20 split between training and test data for validation. 
Model performance is evaluated using classification accuracy for the collider viability branch and mean squared error (MSE) on the log-scale for 
the $\chi^2$ regression branch. After training, predicted $\chi^2$ values are transformed back via 
$\chi^2_{\mathrm{pred}} = \exp(y_{\mathrm{reg}}) - 1$, and a root-mean-square error (RMSE) 
is computed in the original $\chi^2$ domain for interpretability.

\section{Results and Discussion}
\label{sec:results}

\begin{figure}
    \centering
    \includegraphics[width=0.9\linewidth]{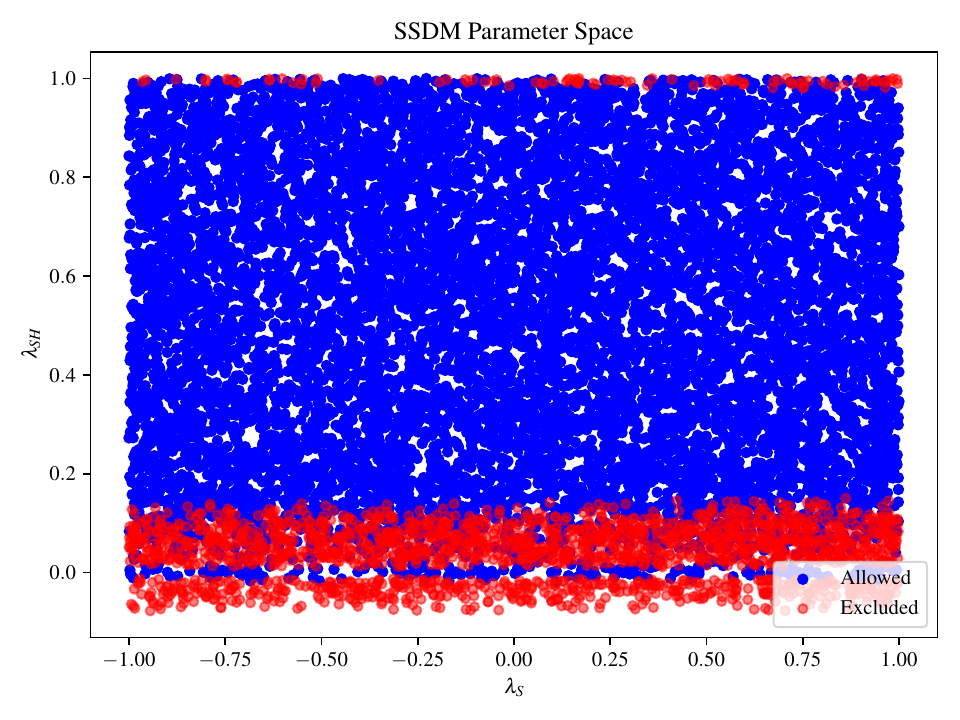}
\caption{Sampled SSDM parameter space showing allowed (blue) and excluded (red) points in the $(\lambda_S,\, \lambda_{SH})$ plane. Exclusion concentrates at small $\lambda_{SH}$ values.}
    \label{fig:ssdm_param}
\end{figure}

\begin{figure}[!ht]
  \centering
  \includegraphics[width=\linewidth]{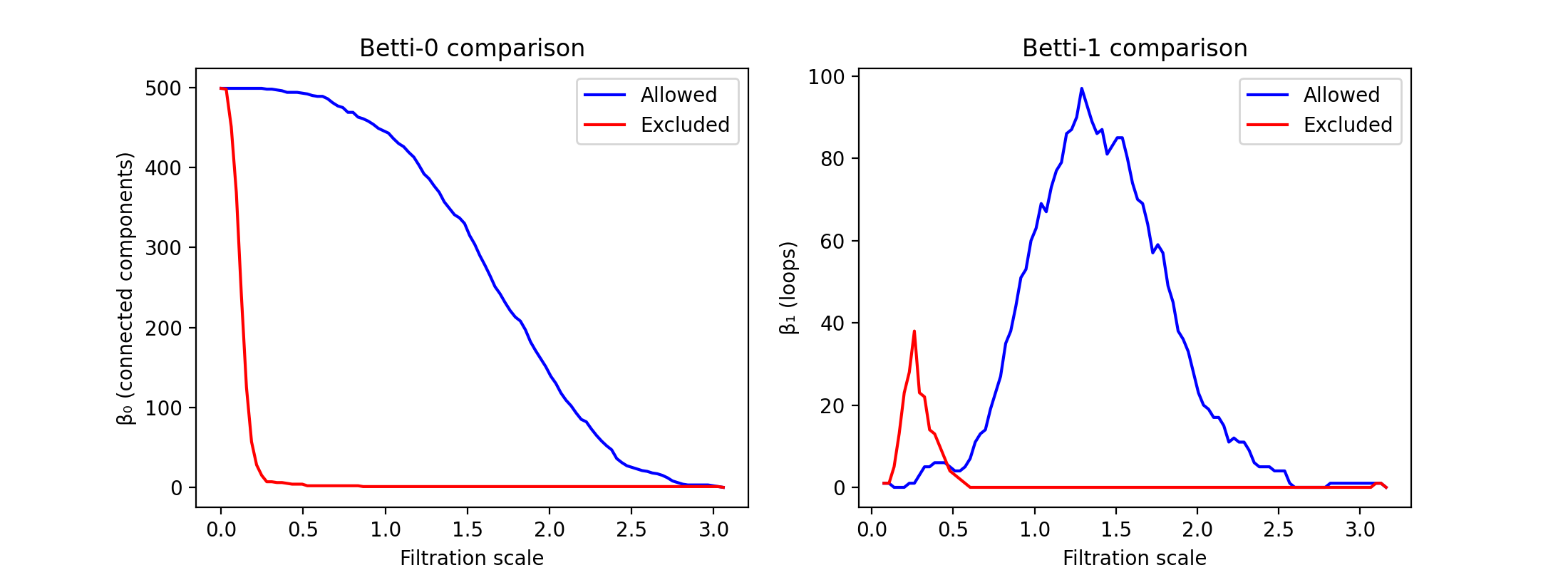}
  \caption{Betti curves for the SSDM parameter space.}
  \label{fig:ssdm-betti}
\end{figure}

\begin{figure*}[!ht]
  \centering
    \includegraphics[width=\linewidth]{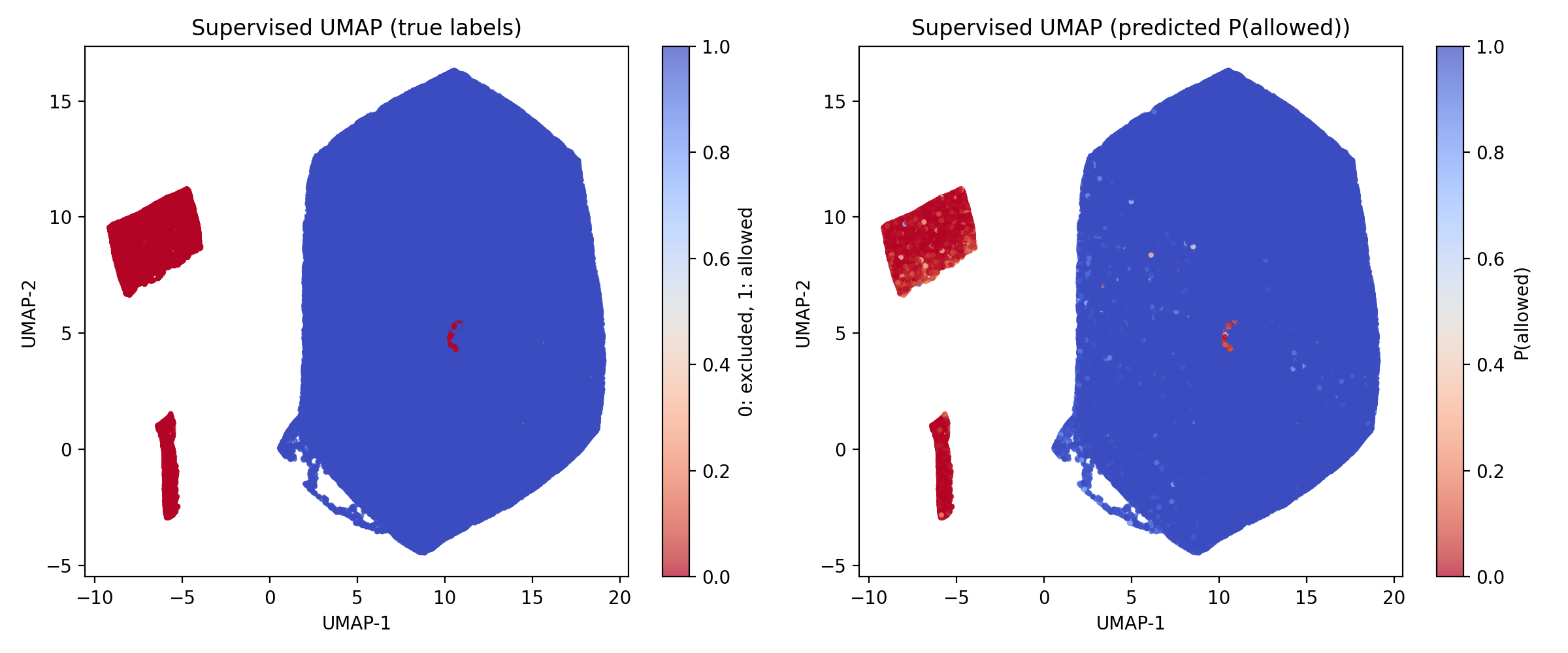}
  \caption{Supervised UMAP projection for the SSDM parameter space. Left: true labels and Right: predicted probability for allowed label.}
  \label{fig:ssdm-umap}
\end{figure*}
\subsection{SSDM parameter space}

As shown in Figure~\ref{fig:ssdm_param}, the sampled Scalar Singlet Dark Matter (SSDM) 
parameter space exhibits a clear separation between collider-allowed and excluded regions 
in the $(\lambda_S,\, \lambda_{SH})$ plane. The distribution indicates that most parameter 
points remain viable across a broad range of the singlet self-coupling $\lambda_S$, whereas 
exclusion predominantly occurs at small values of the Higgs-portal coupling $\lambda_{SH}$. 
This trend reflects the underlying phenomenology: for weak portal interactions, the singlet 
scalar couples too feebly to the Higgs sector, leading to suppressed production cross sections 
and insufficient annihilation rates to satisfy the Higgs signal strength 
constraints. Conversely, moderately larger values of $\lambda_{SH}$ enhance portal-mediated 
interactions, allowing a broader region of parameter space consistent with experimental 
bounds. We also observe overlapping red and blue points, which correspond to different values of the $m_S^2$ parameter.

Such scatter plots provide an intuitive visualization of the viable landscape and 
serve as valuable training targets for the multi-task neural surrogate, which learns to 
approximate the nonlinear decision boundary between allowed and excluded regions while 
simultaneously predicting the associated $\chi^2$ goodness-of-fit values.

Figure~\ref{fig:ssdm-betti} shows the Betti-0 and Betti-1 curves obtained by computing persistent homology from a random sample of 500 allowed and 500 excluded points. The Betti-0 curve ($\beta_0$) quantifies the number of connected components, 
while the Betti-1 curve ($\beta_1$) captures the number of one-dimensional loops across increasing filtration scales. The allowed region exhibits a gradual decay in $\beta_0$ and a pronounced $\beta_1$ peak around intermediate scales, indicating a complex and nontrivial topology. In contrast, the excluded points rapidly collapse into a single connected component with negligible $\beta_1$ structure, consistent with a more compact and topologically simple distribution.

Figure~\ref{fig:ssdm-umap} illustrates the supervised UMAP projections of the SSDM parameter space. The left panel shows the embedding colored by the true class labels, clearly delineating the regions corresponding to allowed (blue) and excluded (red) parameter configurations. The right panel presents the same embedding but colored according to the model’s predicted probability of the allowed class, P(allowed). The close correspondence between the two panels demonstrates that the supervised UMAP successfully preserves the class structure in the low-dimensional manifold, indicating that the machine learning model captures the underlying separation of allowed and excluded regions in the parameter landscape.

\begin{figure}[!ht]
  \centering
    \includegraphics[width=\linewidth]{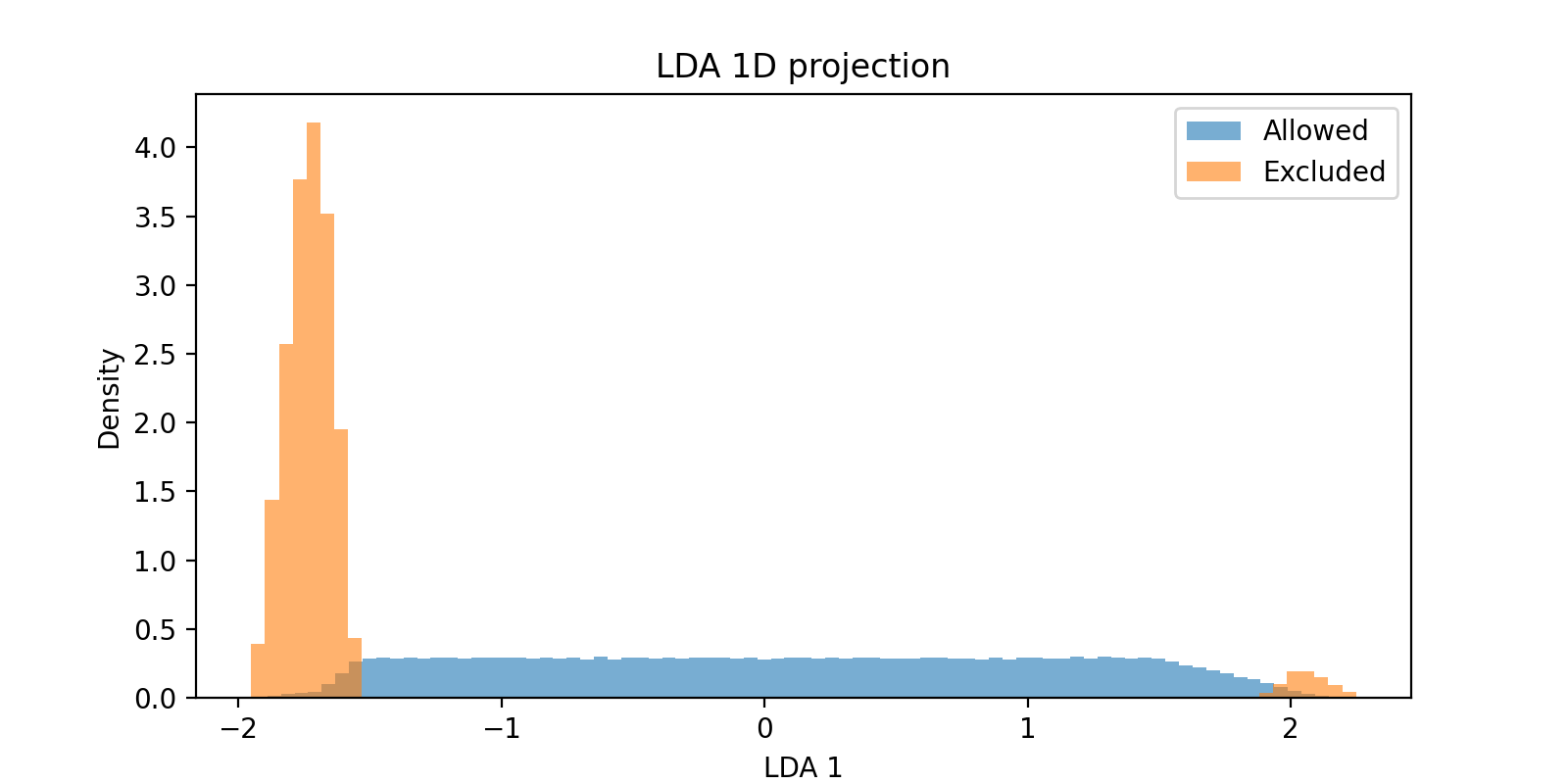}
  \caption{Supervised LDA projection for the SSDM parameter space.}
  \label{fig:ssdm-lda}
\end{figure}

\begin{figure*}
    \centering
    \includegraphics[width=0.45\linewidth]{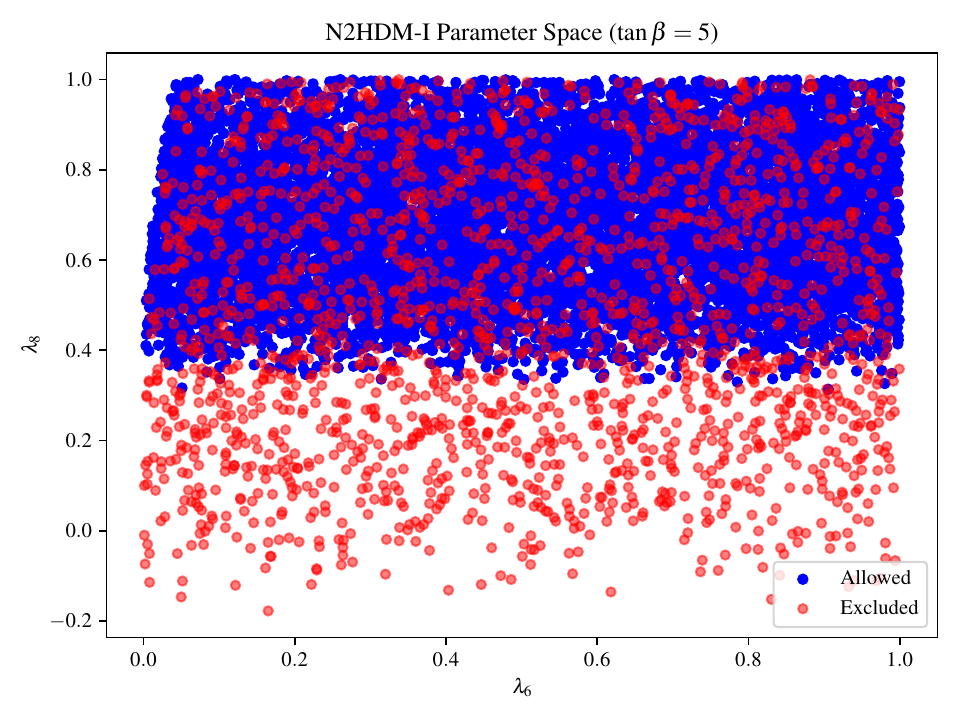}
    \includegraphics[width=0.45\linewidth]{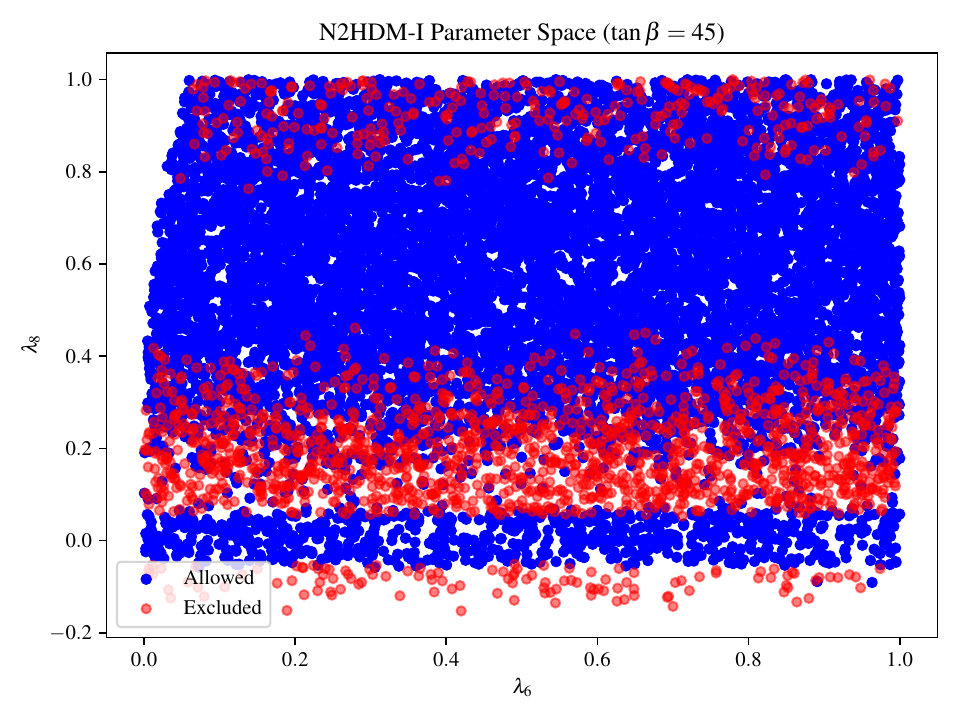}
\caption{Sampled N2HDM parameter space showing allowed (blue) and excluded (red) points in the $(\lambda_6,\, \lambda_8)$ plane.}
    \label{fig:n2hdm_param}
\end{figure*}

Figure~\ref{fig:ssdm-lda} shows the one-dimensional Linear Discriminant Analysis (LDA) projection of the SSDM parameter space, highlighting the class density distributions for allowed and excluded regions. The two classes form well-separated peaks along the LDA axis, indicating that the discriminant component effectively captures the linear boundary between the parameter regimes. This separation confirms that the underlying feature space supports a strong linear discriminative structure between the allowed and excluded configurations.

Finally, using the previously described ML framework, we obtain an allowed–excluded classification accuracy of 99.72\%, accompanied by a corresponding mean squared error in $\chi^{2}$ fit of 20.13.

\subsection{N2HDM Type I parameter space}

Figure~\ref{fig:n2hdm_param} illustrates the sampled parameter space of the Type-I N2HDM in the $(\lambda_6, \lambda_8)$ plane for two representative values of $\tan\beta = 5$ (left) and $\tan\beta = 45$ (right). The blue points correspond to parameter configurations that satisfy all theoretical and experimental constraints, while the red points represent those excluded by the Higgs search at colliders. As described by the scalar potential in Eq. \ref{eq:pot-n2hdm}, the parameters $\lambda_6$ and $\lambda_8$ govern the interactions between the singlet scalar $S$ and the doublet fields $(\Phi_1, \Phi_2)$, thereby influencing the stability and allowed vacuum structure of the model. The dependence on $\tan\beta$, the ratio of the vacuum expectation values of the two doublets, plays a crucial role in determining the allowed regions. At higher $\tan\beta$, the fermionic couplings are suppressed due to the Yukawa structure of the Type-I scenario, which relaxes collider constraints and broadens the viable parameter space. The figure thus captures how varying $\tan\beta$ modulates the interplay between scalar couplings and phenomenological limits, revealing regions where singlet-enriched Higgs states remain consistent with current experimental bounds.

\begin{figure}[!ht]
  \centering
  \includegraphics[width=\linewidth]{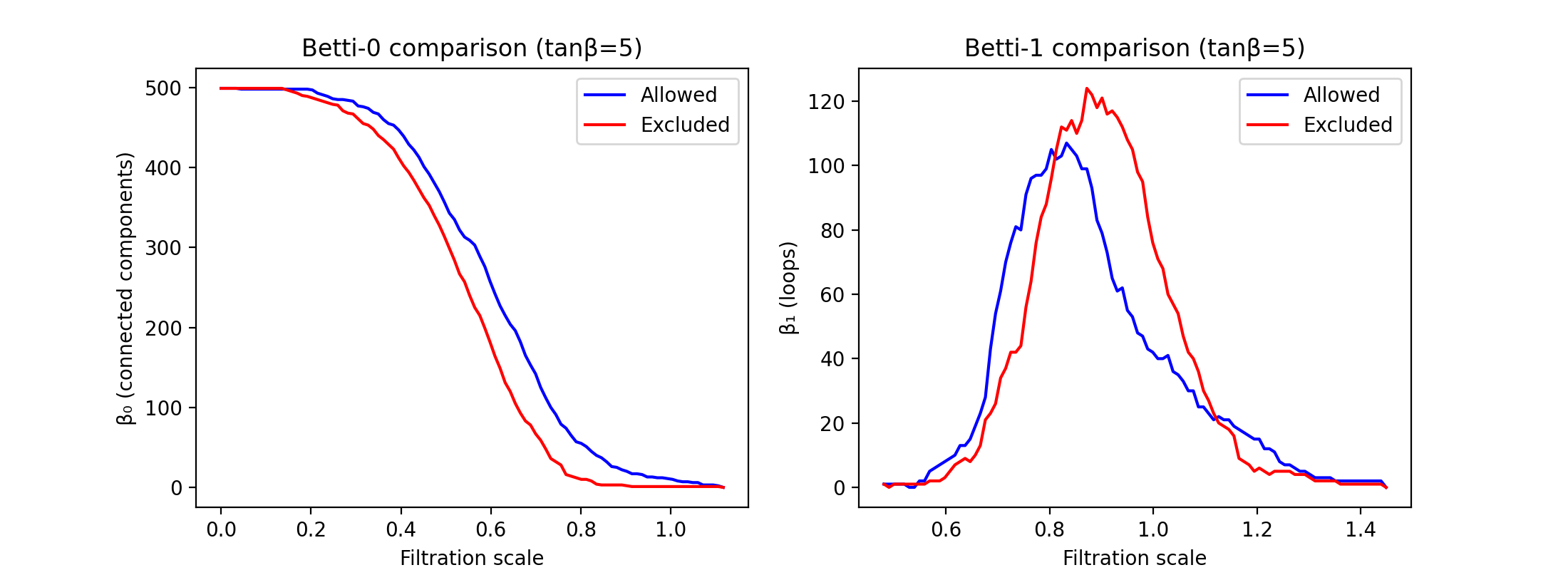}
    \includegraphics[width=\linewidth]{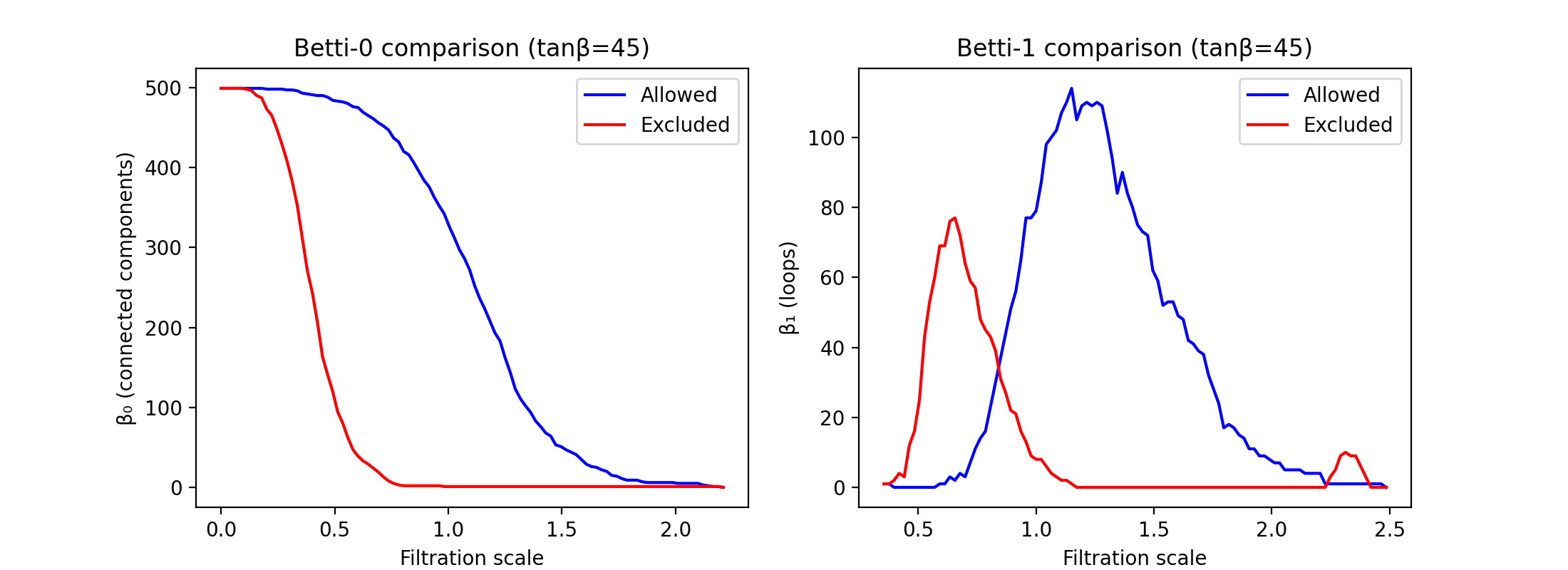}
  \caption{Betti curves for the N2HDM parameter space.}
  \label{fig:ndhdmI-betti}
\end{figure}

Figure~\ref{fig:ndhdmI-betti} presents the Betti curves for the N2HDM Type I parameter space, comparing the topological features of allowed (blue) and excluded (red) regions for two representative values of $\tan\beta$. The first row corresponds to $\tan\beta = 5$, while the second row shows results for $\tan\beta = 45$. The Betti-0 curves capture the evolution of connected components as the filtration scale increases, indicating how rapidly the parameter space becomes topologically connected. The slower decay of the Betti-0 curve for the allowed region suggests a more dispersed and structured connectivity compared to the excluded one. The Betti-1 curves represent the emergence of loops within the parameter space, revealing distinct topological patterns between the two classes. For larger $\tan\beta$, the difference between allowed and excluded regions becomes more pronounced, with the allowed configurations exhibiting richer topological persistence. This signifies a broader and more complex viable region in the parameter landscape.

\begin{figure}[!ht]
  \centering
  \includegraphics[width=\linewidth]{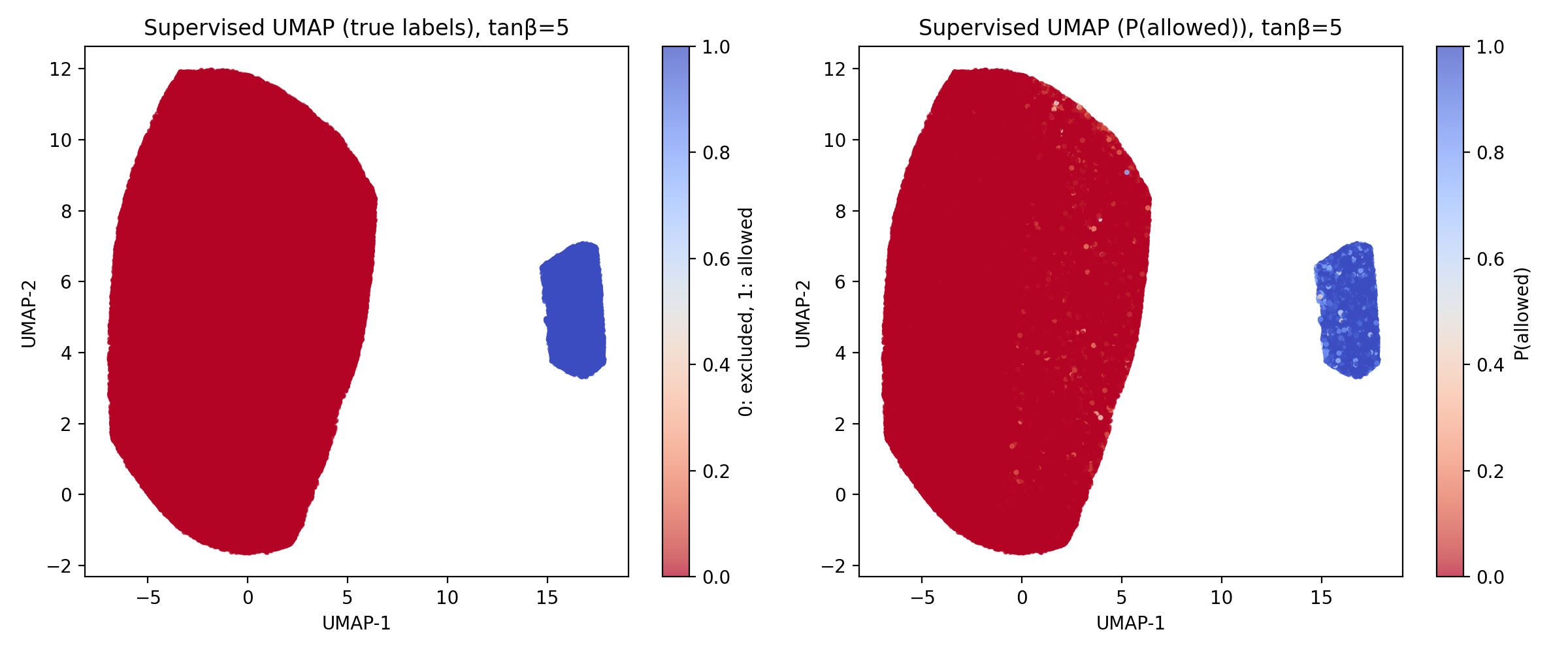}
    \includegraphics[width=\linewidth]{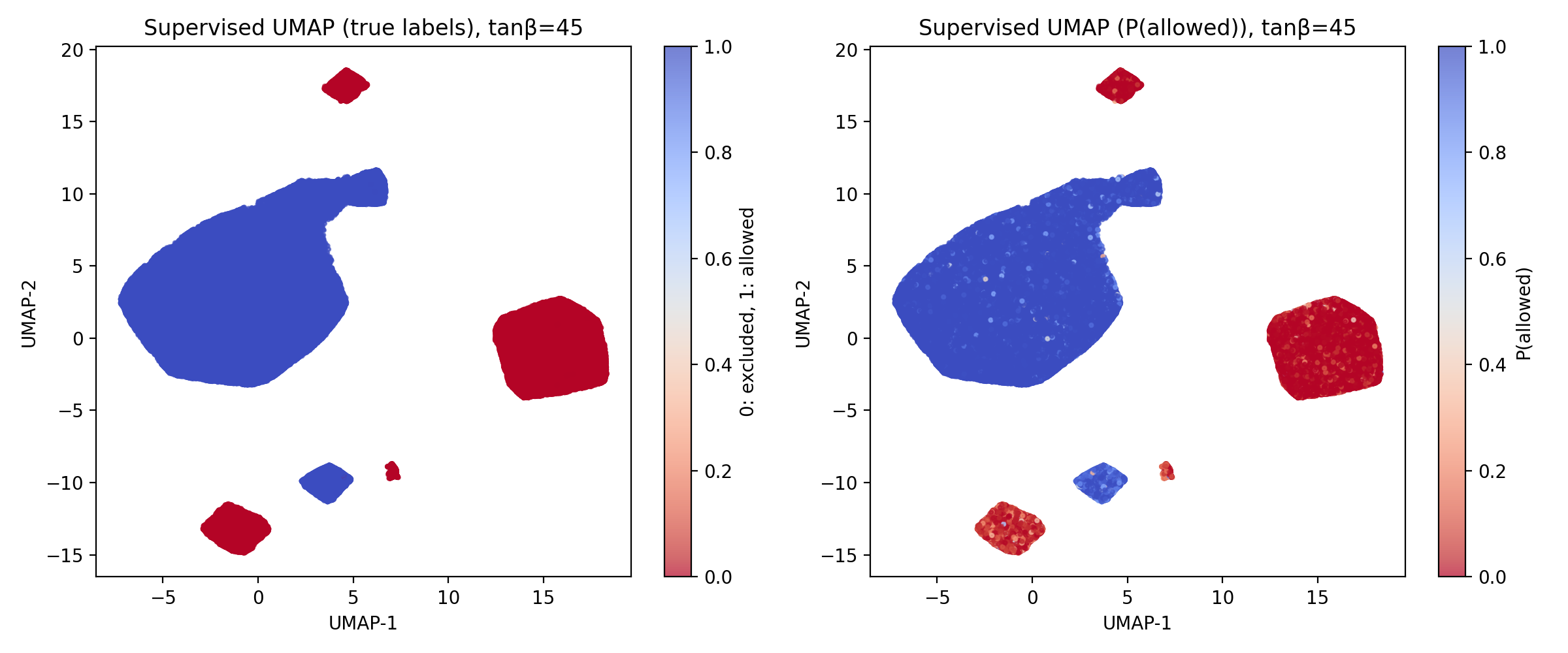}
  \caption{UMAP profile for the N2HDM parameter space.}
  \label{fig:n2hdmI-umap}
\end{figure}

Figure~\ref{fig:n2hdmI-umap} shows the supervised UMAP projections of the N2HDM Type I parameter space for two representative values of $\tan\beta$, with $\tan\beta = 5$ in the first row and $\tan\beta = 45$ in the second. The left panels display the embeddings colored by the true class labels, distinguishing the allowed (blue) and excluded (red) regions, while the right panels represent the corresponding predicted probabilities of the allowed class, P(allowed). For $\tan\beta = 5$, the separation between allowed and excluded regions is sharp and well localized, indicating strong discriminability. As $\tan\beta$ increases to 45, the structure becomes more complex, with the allowed regions exhibiting broader and more intricate connectivity, reflecting the expansion of viable configurations in parameter space. The close correspondence between the true and predicted embeddings further confirms that the supervised UMAP efficiently captures the non-linear decision boundaries learned by the model.

\begin{figure}[!ht]
  \centering
  \includegraphics[width=\linewidth]{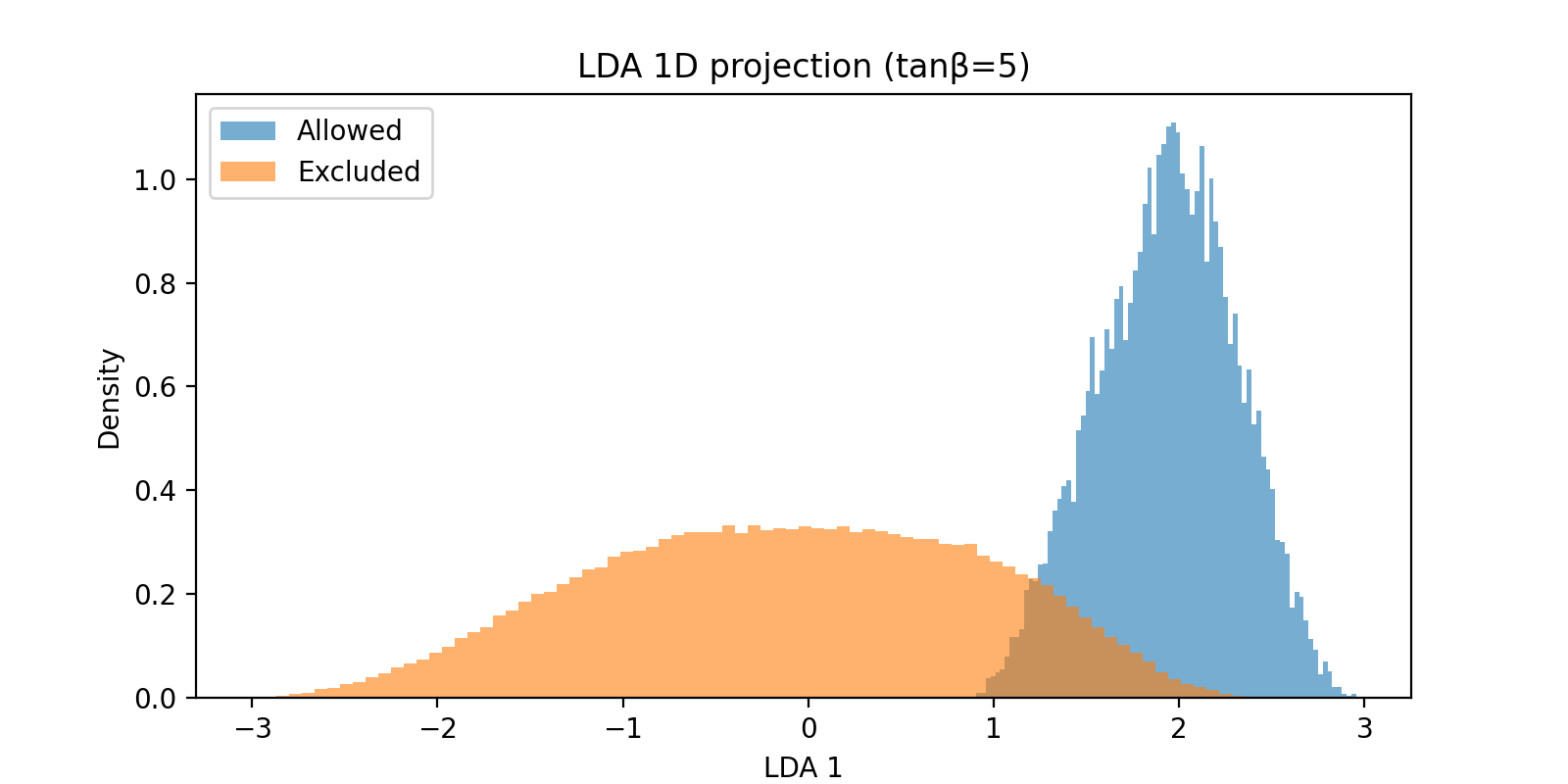}
    \includegraphics[width=\linewidth]{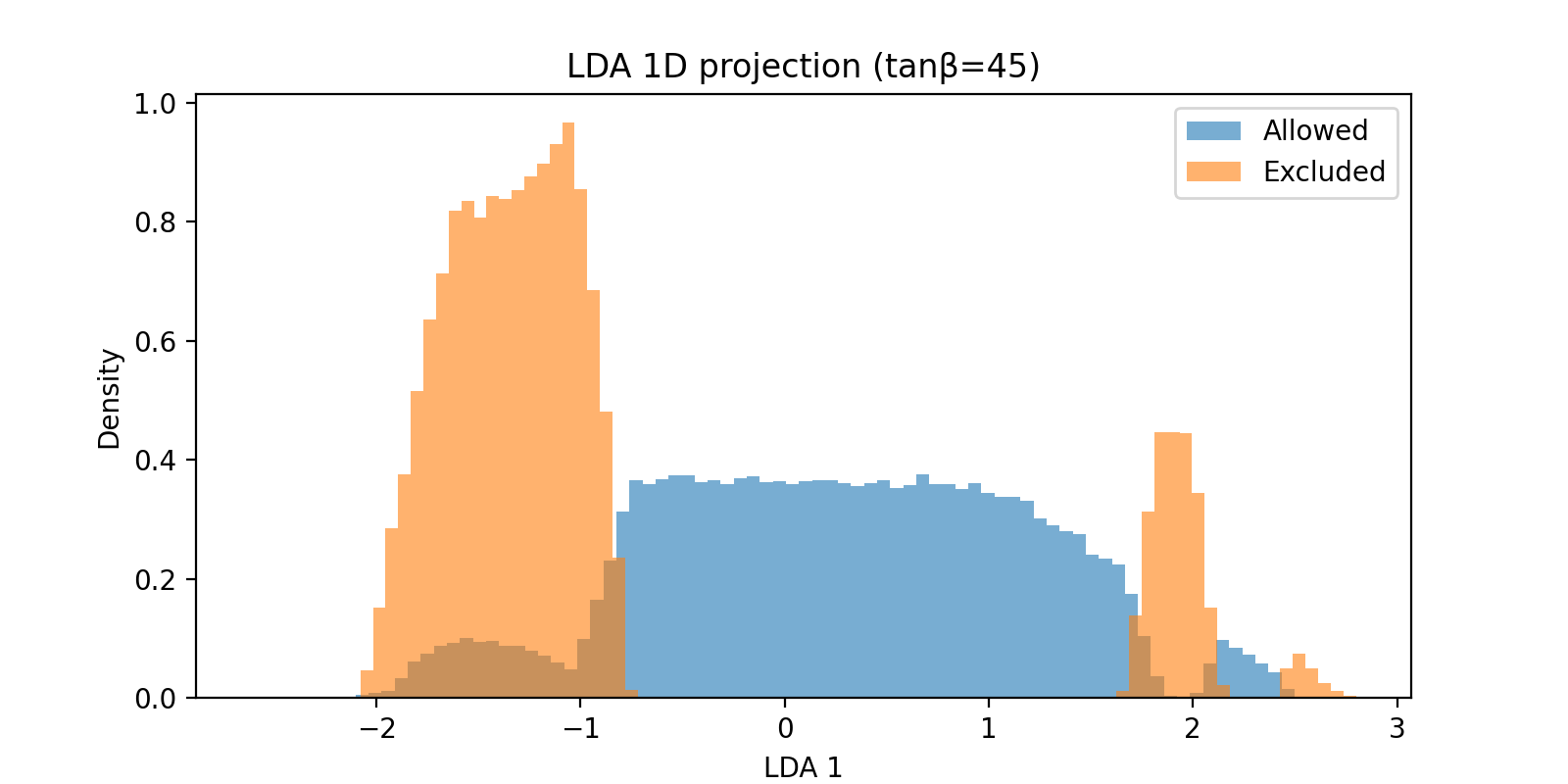}
  \caption{LDA profile for the N2HDM parameter space.}
  \label{fig:n2hdmI-lda}
\end{figure}

Figure~\ref{fig:n2hdmI-lda} presents the Linear Discriminant Analysis (LDA) projections for the N2HDM Type I parameter space, comparing the class density distributions for $\tan\beta = 5$ (top) and $\tan\beta = 45$ (bottom). The LDA axis represents the optimal linear combination of parameters that maximizes the separation between allowed (blue) and excluded (orange) regions. For $\tan\beta = 5$, the two distributions are well separated with minimal overlap, indicating strong linear discriminability between the classes. At higher $\tan\beta = 45$, however, the allowed and excluded regions exhibit partial overlap and a more complex distribution, suggesting a nonlinear mixing of features in the high-dimensional parameter space. This transition highlights how increasing $\tan\beta$ alters the geometry of the viable region, reducing the effectiveness of linear separation and motivating the use of nonlinear embedding techniques for accurate classification.

Using the previously described machine learning framework, we achieve a binary classification accuracy of 99.78\%, with a corresponding mean squared error in the $\chi^{2}$ fit of 20.40 for $\tan\beta = 5$.
Similarly, for $\tan\beta = 45$, the framework attains an accuracy of 99.46\% and a mean squared error in the $\chi^{2}$ fit of 28.03.

\section{Conclusion}
\label{sec:conclusion}

In this study, we employ a combination of topological, manifold learning, and statistical learning methods to investigate the structure of the parameter space in extended Higgs sector models, with a focus on the SSDM and N2HDM frameworks. Through supervised UMAP and LDA projections, we demonstrate that the allowed and excluded regions can be effectively separated, revealing clear geometric and probabilistic boundaries in the multidimensional landscape. The high classification accuracy of greater than $99\%$ and the low mean squared error in $\chi^2$ fit confirm that the machine learning framework reliably captures the intricate correlations among the scalar couplings and potential parameters constrained by theoretical consistency and collider bounds.

The analysis of Betti curves further provides a topological perspective on the morphology of the viable parameter space. Distinct trends in Betti-0 and Betti-1 profiles between allowed and excluded parameter points indicate differences in the connectivity and the presence of complex topological features within the model’s parameter space. In the case of the N2HDM models, the observed evolution of parameter-space geometry with increasing $\tan\beta$ illustrates how phenomenological constraints shape the viable domains, especially in singlet-enriched Higgs extensions.

Taken together, these results emphasize the power of combining machine learning and topological data analysis in probing high-dimensional theoretical models. Though this work emphasises the BSM Higgs sector, the framework can be extended to the complex parameter space of any model with tens of parameters. Such a framework not only enhances interpretability in model selection and constraint visualization but also establishes a foundation for systematic exploration of BSM scenarios where traditional scans over parameter space become computationally prohibitive.

\bibliographystyle{unsrtnat}
\bibliography{ref}

\end{document}